# Real Space Characterization of Nonlinear Hall Effect in Confined Directions

*Sheng Luo[1], Chuang-Han Hsu[1,2], Guoqing Chang[3], Arun Bansil[4], Hsin Lin[2\*], Gengchiau Liang[1\*]*

[1] Department of Electrical and Computer Engineering, National University of Singapore, Singapore 117576

[2] Institute of Physics, Academia Sinica, Taipei, 11529 Taiwan

[3] Division of Physics and Applied Physics, School of Physical and Mathematical Sciences, Nanyang Technological University, Singapore, Singapore

[4] Department of Physics, Northeastern University, Boston, MA 02115, USA

ABSTRACT: The nonlinear Hall effect (NLHE) is a phenomenon which could produce a transverse Hall voltage in a time-reversal-invariant material. Here, we report the real space characterizations of NLHE evaluated through quantum transport in $TaIrTe_4$ nanoribbon without the explicit Berry curvature dipole (BCD) information. We first characterize the NLHE in both transverse confined directions in global-level measurement. The impact of quantum confinement in NLHE is evaluated by adjusting the width of nanoribbons. Then, the probing area is trimmed to the atomic scale to evaluate the local texture, where we discover its unique patterns among the probed atomic groups for the first time. The analysis of charge distribution reveals the connections between NLHE's local patterns and its non-centrosymmetric nature, rendering nearly an order of Hall voltage enhancement through probe positioning. Our work paves the way to expand the range of NLHE study and unveil its physics in more versatile material systems.

**Introduction**

The Hall effects have been one of the crucial topics in condensed-matter physics [1-2] and form the basis of fundamental research of material properties and practical applications. In recent years, the nonlinear Hall effect (NLHE) in the time-reversal invariant system is proposed [3], and subsequently verified in various two-dimensional materials (2D materials) through experiments, such as $WTe_2$ and $TaIrTe_4$ [4-7]. The NLHE demonstrates a second-order response from the applied electric field in a time-reversal invariant system, different from the other Hall effects occurring in the linear regime, such as the anomalous Hall effect (AHE) [2]. The NLHE sparks intense research interests, and its second-order response feature renders several promising applications such as radiofrequency rectifier [7-8], memory [9], and energy-harvesting devices [10], etc.

The Berry-curvature dipole (BCD) induced NLHE is considered as an important mechanism in the field of NLHE study [3,11-12]. In this mechanism, the BCD could act as an effective magnetic field to allow the chiral movement of electrons and induce anomalous velocity to the system. In a time-reversal invariant system, the non-zero BCD could be induced by broken inversion symmetry. The corresponding second-order Hall conductivity is then characterized through semi-classical Boltzmann transport equations (BTE) [3]. Besides, different from the scattering-free intrinsic anomalous [2] and spin Hall effects [13], the BCD-induced NLHE is scattering-dependent, and it is different from the NLHE induced by the extrinsic mechanisms, such as disorder-induced scattering [11]. The BCD is an inherent property of the Hamiltonians [3], and it requires at least two periodic dimensions in $k$-space [14]. Due to the presence of confinement in some materials, such as the out-of-plane direction in 2D material, the BCD could not be extracted from band structures. This restricts the scenarios where BCD could be implemented to characterize the NLHE, especially as several 2D materials, including WTe$_2$ [4-6], TaIrTe$_4$ [7], MoTe$_2$ [15], are important platforms for NLHE research. Therefore, it is desirable to further develop the characterization scheme of NLHE for more versatile material systems without the explicitly defined BCD.

To solve the constraints of NLHE's characterizations in confined systems, the quantum transport simulation framework including non-equilibrium Green's function (NEGF) formalism [16] with tight-binding Hamiltonian models [4,17-18] could be considered as a solution. The NEGF formalism does not require the explicit presence of BCD in the derivation of nonlinear Hall current, but instead directly provides the overall transverse Hall current or voltage. As the inherent property of the Hamiltonians, the adiabatic contributions of the BCD could be directly involved in the NEGF scheme. Besides, the tight-binding (TB) model extracted from *ab initio* with non-zero BCD [7] could directly provide real-space information. Combining with the NEGF formalism, the carrier transport theorem could reveal not only the 2nd-order response of NLHE, but also the behaviors of NLHE's mechanisms in real space, a new aspect different from the *k*-space dependent BCD could provide.

In this work, we report the theoretical characterization of BCD-induced NLHE in bi-layer TaIrTe$_4$ nanoribbons with the width of a few nanometers under direct current (DC) conditions. We first verify the presence of NLHE in TaIrTe$_4$ nanoribbons in both confined transverse directions in global-level measurement [as shown in Fig. 1(a) and (b)]. We consider the NLHE in both transverse directions due to the remnants of BCD involved in carrier transport states under quantum confinement. The scaling trend of $V_{NLHE}$ with ribbon width further demonstrate the impact of quantum confinement through the analysis of band structure. Subsequently, the probing area is further reduced to the atomic scale to investigate the real-space textures of NLHE on local level [as shown in Fig. 1(c)]. The unique real-space patterns are observed, including the polarity aversions of Hall voltage among neighboring atomic groups. We further discuss the origin of these patterns as the results of the non-centrosymmetric charge distribution of in TaIrTe$_4$ nanoribbon, derived from its electronic structure. Furthermore, the pattern provides the opportunity to boost NLHE response by nearly an order through the optimization of the probe's positions, offering another method to adjust the response of NLHE in device applications. Our works outlines a novel approach to unveil the nature of NLHE in versatile material structures with real-space aspects, providing the unique insight for device applications.

**Characterization of NLHE in global-level probing**

To verify the presence of NLHE in TaIrTe$_4$ nanoribbon though NEGF scheme, voltage measurement through large area probing on nanoribbon is performed to characterize NLHE in both transverse directions, as shown in Fig. 1 (a) and (b). The "global" probing configurations aim to approach the experimental set-up of large area metal contact in voltage measurement, and further examine if the extracted nonlinear Hall voltage ($V_{NLHE}$) in 1D ribbon is indeed the NLHE. According to the BCD component $D_{xz}$ (Fig. 2a), a change of polarity could be observed at different chemical potentials, and it could be served as a criterion to verify the origin of in-plane $V_{NLHE}$. Therefore, two Hall voltage measurements are designed in transport windows centering at 0.0 and 0.1 eV, respectively. The system is biased ranging from 0 to 10 mV, ensuring the same BCD polarity within the transport window. Furthermore, a potential $U(x)$ of first-order approximation is applied to the system under small bias is applied to the system, introducing a non-rigid scattering source. It is defined as $U(x) = \frac{\mu_1 - \mu_2}{L} x$, where $\mu_1$ and $\mu_2$ are the fermi-level defined in the lead regions and *L* is the length of the scattering region (2.3 nm). Both the bulk and edge states of nanoribbons are involved in carrier transport as shown in Fig. 2(b) for both evaluated transport windows. Through the calculations of the probed chemical potential under different applied bias, the results in Fig. 2(c) and (d) demonstrate the presence of both in-plane and out-of-plane NLHE. By applying both forward and reverse bias to the system, the extracted NLHE induced electric field ($E_H$) remain the same without polarity aversion. This consolidates the Hamiltonian model as it has properly constructed TRS and excludes the presence of anomalous Hall effect (AHE), while the 2nd-order components dominate the carrier transport.

For the origins of the observed $E_H$, the in-plane Hall voltage indeed shows the reversed polarity under different transport windows centered at 0.0 and 0.1 eV (Fig.2c), in accordance with the BCD polarity (Fig. 2a). The match of the polarity between the BCD and the NEGF-derived $E_H$ demonstrates the BCD-induced NLHE has the dominant contributions to the NLHE observed in the system. To further consolidate its origin, the NLHE evaluated through BTE [3] (see Methods for detail) is carried out to demonstrate the general trend of second-order relation and Hall voltage polarity. The BCD induced $E_H$ calculated through BTE are shown in Fig. 2(c) as the red and blue dashed lines, representing the NLHE in 2D TaIrTe4 sheet, where $E_H$ is proportional to BCD component $\mathcal{D}_{xz}$. Both results show similar second-order relation between the *E*H and the longitudinal applied field. The same Hall voltage polarity corresponds to the BCD polarity in the transport window (Fig. 2a). Both observations consolidate the NLHE observed in the NEGF-derived results in a 1D nanoribbon system. The main difference between the BTE-/NEGF-derived results stems from the explicit BCD implemented in the BTE-derived second-order conductivity, as it corresponds to 2D bi-layer TaIrTe4. On the contrary, the NEGF-based results derived from 1D nanoribbon. Such variation is induced by quantum confinement as the shifting of the transport states changes the nonlinear Hall current/conductance in transverse direction. To further investigate the impact of quantum confinement, the scaling trend of in-plane $E_H$ with nanoribbon width is evaluated with various ribbon width scaling from 2.5 nm to 6.9 nm, and the corresponding band structures are shown in Fig. 3(a). The in-plane $E_H$ are extracted following the configurations in Fig. 1(b) in transport window centered at 0.2 eV. The scaling trend in Fig. 3(b) demonstrates the extracted $E_H$ positively correlated to the ribbon width under the same applied bias, with BCD induced $E_H$ under 2D structure exceeding nanoribbon's results. The evolutions of band structure from nanoribbon to 2D sheet (Fig. 3a) show as the system reduced from 2D to 1D, the quantum confinement effect induces both band folding and sub-band energy shifting, leading to the reduction of electron states within the transport window, alongside with the remanent BCD participated in carrier transport. The results reflect the nature of NLHE while providing the insight for the adjustment of the NLHE response through material engineering.

Notably, the 2nd-order nonlinear relations between $E_H$ and $E_x$ could be observed in both transverse confined directions. Although the corresponding BCD could not be explicitly obtained due to the confinement, the out-of-plane NLHE (Fig. 2d) could be characterized under the NEGF framework. Comparing to the in-plane components (from $D_{xz}$), the out-of-plane NLHE could not be directly evaluated through BTE. Besides, for bi-layer $Td$-TaIrTe4 nanoribbon, only the in-plane mirror line exists (see Supplementary Figure S2) to constrain the in-plane NLHE. This is different from its bulk counterpart where the out-of-plane 2[1] axis suppresses the out-of-plane NLHE components [19]. The symmetry of bi-layer TaIrTe4 allows the observation of out-of-plane NLHE response in Fig. 2(d) on ribbon surface. The origins of the NLHE observed in both confined transverse directions could be the remnant BCD induced from band folding under quantum confinement, as the system reduces to lower dimension [20], such as the 1D nanoribbon in this work. Furthermore, the mirror symmetry still involves in the carrier transport, as zero Hall voltage could be detected when bias applied along in-plane mirror axis (see Fig. S2 in Supplementary Information), demonstrating that the NLHE from remnant BCD upholds the mirror symmetry constraints.

**Local textures of NLHE from miniaturized probing area**

The real-space characterization of NLHE through the probed chemical potential allow the revelations of local texture of NLHE, in which the $k$-space BCD requires extra procedures to project its characteristics to the real space through BTE [3]. The local effects of $V_{NLHE}$ are evaluated in trimmed probing areas, which reduced to a few atoms in measurement. A cross-section of the nanoribbon at the half of the channel length ($L/2$) is selected to study the local effects (Fig. 1c), with probe scanning around the edges of the cross-section as shown in Fig. 4(a) and extracting the potential from the outermost atoms. For top/bottom edges the probe is connected to two $Te$ atoms in each measurement, and to either one $Te$ or $Ta$ atom when scanning along the left/right edges. Each probe position is labelled regarding the edge (*A, B, C, D*) and scan sequence (1,2,3,4…), as shown in Fig. 4(a).

The probe results are shown in Fig. 4(b) under the applied bias of 10 mV. The probed potential of the atom group "$A_1$" in Fig. 4(b) is used as the reference point for the following results. The probed potentials in each two opposite edges (top/bottom and left/right) differ, revealing the presence of non-zero transverse Hall voltage in the atomic level. To further examine the nonlinearity of the Hall voltage characteristics, a series of longitudinal bias ranging from 0 mV to 10 mV are applied. In Fig. 4(c), the extracted Hall voltages are defined as chemical potential differences between the atomic positions at opposite edges' mirror points, demonstrating nonlinear characteristics' presence in local level.

The probed chemical potential results shown in Fig. 4(b) could produce the maximum $V_{NLHE}$ value up to ~9 mV (using point $A_1$'s value as reference) biased at 10 mV by repositioning the voltage probe between $A_4A_1$ or $A_4D_3$ atom groups instead of the mirror positions (Fig. 4a). This is a value nearly an order higher than $V_{NLHE}$ extracted from the opposite edges as shown in Fig. 4(c) under the same bias. As the probe scan in the positive *y* direction, the polarity aversion among the neighbouring probing positions could be observed (i.e., $V_{A1C1} \to V_{A4C4}$). Different from the polarity aversion of global $V_{NLHE}$ reflecting the in-plane BCD's polarity, these local-level aversions in $V_{AC}$ could not be directly perceived from the *k*-space dependent BCD. To further evaluate the asymmetric and variated Hall voltage behaviors, the charge distributions in the selected cross-section are shown in Fig. 4(d). The non-centrosymmetric distributions of charge centroid within the cross-section could be observed, in accordance with the chemical potential probed in local atom groups. The asymmetrically distributed charge centroids reflect the broken inversion symmetry of the TaIrTe$_4$, and further inducing the aversion of nonlinear Hall voltage's polarity in measurement among opposite edges, as shown in the red lines of Fig. 4(c). Besides, the non-trivial variations of probed nonlinear Hall voltage among neighboring atomic groups are also in accordance with the charge centroid distribution shown in Fig. 4(d), revealing the patterns of BCD projected onto different orbital states along the edges under the quantum confinement.

**Conclusions**

In summary, we have characterized and investigated the NLHE in bi-layer TaIrTe4 nanoribbon, demonstrating the $V_{NLHE}$ in both confined transverse direction in 1D nanoribbon system, and the textures of locally probed Hall voltage in atomic resolution without the explicit BCD information. The scaling trend of NLHE with ribbon width further reveals the impact of quantum confinement effect on NLHE. The miniaturized probing areas also unveil the unique asymmetric variations and polarity aversions of the Hall voltage among the neighboring atomic groups, revealing the real-space patterns for the first time. The global level probing approaches the k-space BCD characteristics, while the local level measurement reveals the real-space projections of BCD among different atomic groups. Furthermore, the microscopic probing of NLHE in confined nanostructures demonstrates that the NEGF formalisms with real-space model could provide the numerical evaluation of the remanent BCD components in constraint dimensions. As NEGF formalism allows direct evaluation of overall Hall voltage without explicit BCD, the research of NLHE could be expanded into more versatile material systems. For instance, the edge states from various materials could be examined for the 2nd-order nonlinearity of Hall voltage in low dimensional system, such as the 2D topological insulator (TI) materials. The real-space characterization of $V_{NLHE}$ also allows the optimization of nonlinear Hall effect-based device's performance, such as enhancing $V_{NLHE}$ response through probe-position design as demonstrated in Fig. 4(b). The overall real-space characterization of NLHE not only allows expanding the NLHE into more versatile material systems, but also paves the way for further adjustment of NLHE in nanoscale device applications.

**Methods**

**Derivation of BCD of bi-layer TaIrTe4**

The DFT-based tight binding (TB) model of bulk and bi-layer $T_d$-TaIrTe4 are derived from the first-principles calculations with the same method applied in the previous studies [7,21]. In this procedure, a bulk TB model of $T_d$-TaIrTe4 is first interpolated from the standard scheme of maximally-localized Wannier functions as implemented in Wannier90 package [22], then the bi-layer $T_d$-TaIrTe4 is constructed by truncating the bulk TB model. Both the 2D sheet and 1D nanoribbon band structure of bi-layer $T_d$-TaIrTe4 could be found in Supplementary Fig. S1. The Berry curvature dipole $D_{ab}$ of bi-layer TaIrTe4 is then calculated via [3],

$$\mathcal{D}_{ab} = \int_k \frac{dk^2}{(2\pi)^2} \sum_n \frac{\partial \varepsilon_n(k)}{\partial k_a} \Omega_n^b(k) \left(-\frac{\partial f_0(k)}{\partial \varepsilon_n(k)}\right). \quad (1)$$

where $\varepsilon_n(k), k_a, f_0,$ and $\Omega_n^b(k)$ are band energy of *n-th* band at k point *k*, a-component of k vector, the Fermi-Dirac function, and b-component of Berry curvature vector (a and b = *x, y, z* in the Cartesian coordinate).

**Hall voltage calculation in NEGF formalism**

The NEGF [16] formalism is implemented for carrier transport calculations under ballistic transport assumption, as the longitudinal TaIrTe4 nanoribbon length scaled down to ~2.3 nm. Semi-infinite boundary conditions have been added to the system, a schematic view of the simulated system is demonstrated in Fig. S4. The surface green's functions are calculated based on the Sancho-Rubio's scheme [23], a standard method accounting semi-infinite boundary conditions in NEGF. The carrier transport is evaluated under the temperature of 0 K.

The voltage probe is placed in transverse surfaces/edges to serve as the third terminal and could be treated as the charge reservoir characterized by the chemical potential *μ*. The local potential could be probed as $\mu$ when the probe current reduces to zero, as both the probe and probing area share the same potential [16,24-25]. It is worth noting that the applied potential $U(x)$ is cancelled out as the probes are placed in the same $x$ coordinate. The corresponding Hall voltage could be obtained through the measurement of local potential $\mu$ in the designated probing areas. The probe configurations only differ in the atoms probe connecting to and the number of atoms probed simultaneously, enabling the real-space characterization of NLHE in both global and local levels. Further discussions regarding the probe parameters and robustness of the probed chemical potential are shown in Supplementary Note 1. For the characterization of the NLHE, the electronic structure of the channel material determines the nonlinear Hall current, which could be evaluated by the NEGF with the modelled voltage probe and leading to the measurement of the corresponding Hall voltage. This allows the direct evaluation of NLHE from electronic structure without the explicit BCD information. The overall response of NEGF should consist of all different orders of Hall current, but due to the presence of TRS and mirror symmetry in bi-layer TaIrTe$_4$ structures, the second-order components dominate the extracted nonlinear Hall voltage. It is worth noting that the recently discovered third-order nonlinear Hall effect has been observed in both thick $T_d$-MoTe$_2$ [26] and TaIrTe$_4$ [19] samples, where both second and third-order responses have similar amplitude. It should be the next non-vanishing response in the NEGF calculations, although the ultra-thin bi-layer TaIrTe$_4$ studied in this work enable the second-order component as the dominant response. This is further discussed in Supplementary Note 2. Furthermore, the non-rigid scattering induced by the applied potential is found to be the main contributing scattering mechanism for the observed NLHE (Fig. S5). Although there could be contributions from other non-rigid scattering mechanisms, they have minor impact in the overall extracted nonlinear Hall voltage. This is further discussed in Supplementary Note 3.

**In-plane NLHE calculations by Boltzmann transport equations**

The nonlinear Hall effect induced by Berry curvature dipole (BCD) could be evaluated through the framework of angular-dependent second-order nonlinear response, well established in the experimental characterizations regarding WTe2 [4,6] and bilayer Graphene [17]. As the bi-layer $T_d$-TaIrTe4 shares the same group symmetry (*Pmn*2$_1$) as $T_d$-WTe2, we could reach the similar angular dependence relation between in-plane applied electric field $E_\parallel$ and perpendicularly probed second-order electric field $E_\perp^{(2)}$ as:

$$\frac{E_\perp^{(2)}}{E_\parallel^2} = \rho_{yy} \chi_{yxx} \tag{2}$$

where $\rho_{yy}$ is the resistivity along the mirror line. To provide a meaningful comparison with the quantum transport simulation, this parameter could be extracted under the ballistic limit through NEGF calculation at 0K as:

$$G_{yy} = \frac{q^2}{h} \cdot \frac{W}{2\pi} \int T(E, k_x) dk_x \tag{3}$$

where $T(E, k_x)$ represents the transmission coefficients under specific energy $E$ according to the selected transport window and transverse mode $k_x$. Integral over the first Brillouin zone and normalized by the transverse width $W$ will further yield the two-dimensional (2D) conductivity $\sigma_{yy}$, which is the reciprocal of the resistivity $\rho_{yy}$. Furthermore, the 2nd-order conductivity $\chi_{yxx}$ in Eq. 4 could be derived with Berry curvature dipole $\mathcal{D}_{xz}$ through [3]:

$$\chi_{yxx} = -\varepsilon_{yzx} \frac{e^3 \tau}{2\hbar^2 (1+i\omega\tau)} D_{xz} \tag{4}$$

where $\varepsilon_{yzx}$ is the antisymmetric tensor, $\omega$ as the applied field's frequency and $\tau$ is the overall relaxation time, which could be expressed under the Matthiessen's rule [27] as:

$$\frac{1}{\tau} = \frac{1}{\tau_B} + \frac{1}{\tau_1} + \frac{1}{\tau_2} + \cdots \tag{5}$$

where $\tau_B$ represents the transit time under the quasi-ballistic conditions, $\{\tau_1, \tau_2, \dots\}$ represent the mean free time of multiple scattering events. As the longitudinal distance of the device $L$ (~ 2.3 nm) is far below the scattering mean free path λ, the overall relaxation time could be approximated by $\tau_B$ as $\tau_B = L/\langle v \rangle$, where average group velocity $\langle v \rangle$ relates to the material's band structure, representing the maximum estimations of the relaxation time. Different electric field applied to the system affects the width of transport window involved in the carrier transport, and thus number of states participated in the transport are affected, which induced the field-dependency on the average group velocity $\langle v \rangle$. Due to the symmetry constraints and system's dimensionality, only the BCD component $D_{xz}$ contributes to the 2nd-order conductivity. Besides, under the low frequency conditions $w\tau \ll 1$, the 2nd-order conductivity could be approximated as $\chi_{yxx} \sim \frac{e^3 \tau}{2\hbar^2} D_{xz}$. By obtaining the $\rho_{yy}$ and $\chi_{yxx}$, the BCD induced Hall electric field under 2D structure could be derived under various applied electric field. It is worth noting that the BTE-derived result represents the estimation of NLHE under the quasi-ballistic condition, which allows it to verify the origin of the NEGF-derived results in Fig. 2c.

## Data availability

The data that support the plots within this paper and the other findings of this study are available from the corresponding author upon reasonable request.

## Code availability

The codes that support the plots within this paper and the other findings of this study are available from the corresponding author upon reasonable request.

## Acknowledgements

This work at the National University of Singapore is supported by MOE-2017- T2-2-114, MOE-2019-T2-2-215, and FRC-A-8000194-01-00. The work at Northeastern University was supported by the Air Force Office of Scientific Research under award number FA9550-20-1-0322, and it benefited from the computational resources of Northeastern University's Advanced Scientific Computation Center (ASCC) and the Discovery Cluster.

## Author contributions
L.H. and L.G. proposed the idea of NLHE project. L.S. conceived and performed the numerical NEGF simulations of the designed systems and analysed the data with L.G. and L.H. C.H. provided the tight-binding model and Berry curvature dipole of the bi-layer TaIrTe$_4$. L.S. wrote the manuscript with input from all other authors.

**Corresponding authors**

Correspondence and requests for materials should be addressed to L.G. (Email: elelg@nus.edu.sg) and L.H. (Email: nilnish@gmail.com).

**Figures**

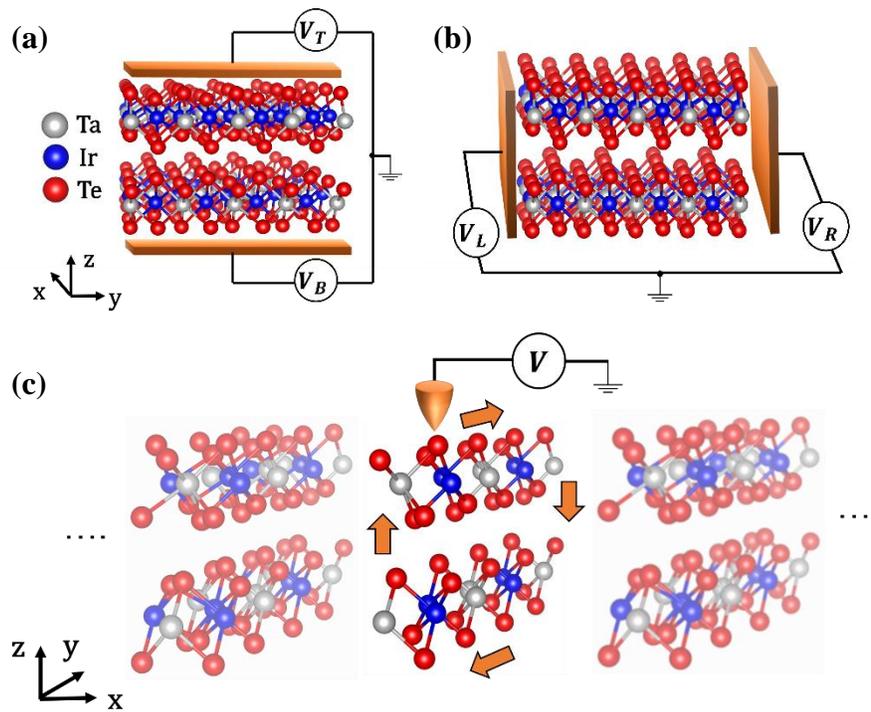

FIG. 1. **Different nonlinear Hall effect measure schemes.** (a) and (b) Probe covering the entire ribbon's surfaces. The longitudinal transport direction is at *x* direction, while the transverse voltage was probed at *z* and *y* direction. (a) is for top/bot surface probe (measured as $V_T$ and $V_B$) and (b) is for left/right edge probe (measured as $V_L$ and $V_R$) to measure the out-of-plane and in-plane Hall voltage, respectively. Each probe's measurement ($V_T, V_B, V_L, V_R$) was conducted separately. (c) Miniaturized probe with resolution of 1-2 atoms. The reduced-size probe aims to characterize the real-space texture of NLHE.

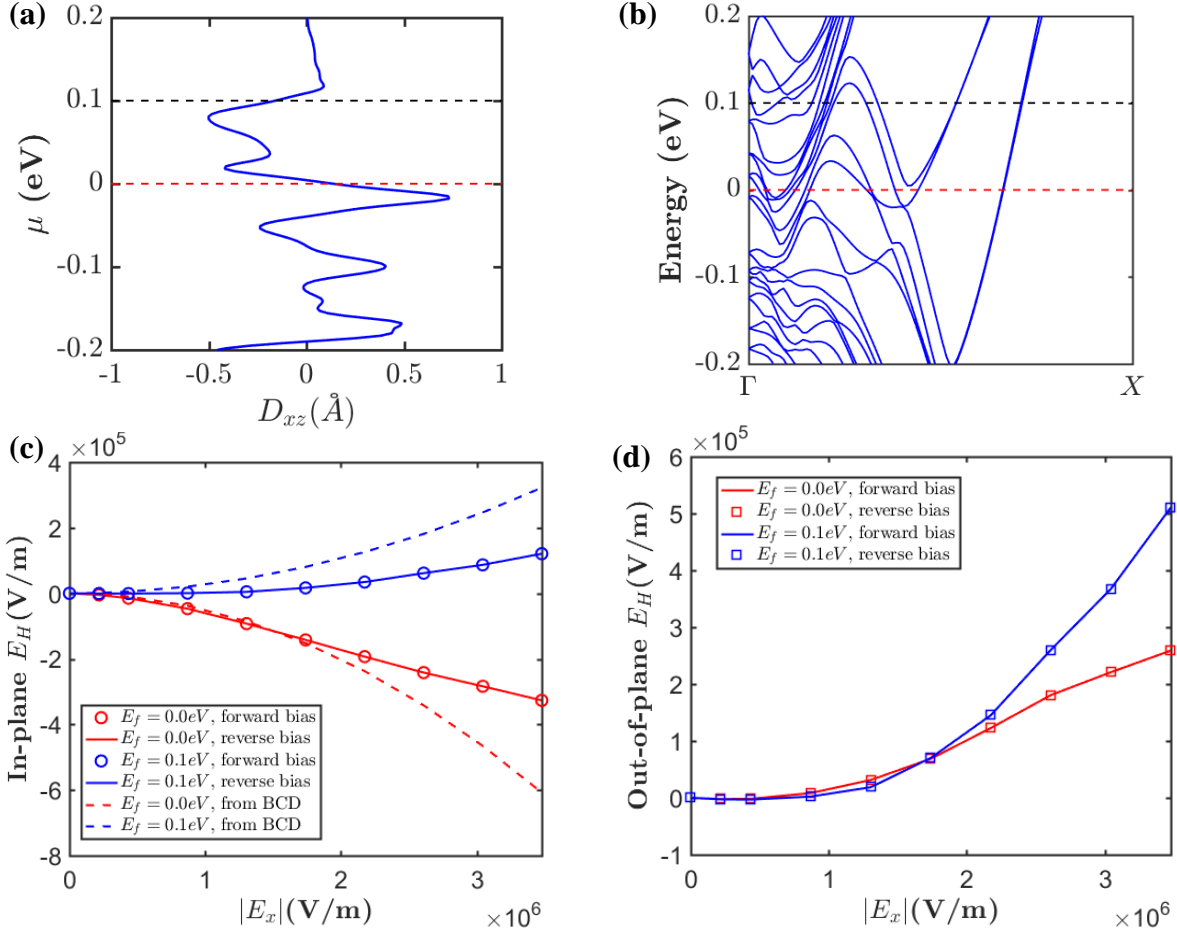

FIG. 2. **Nonlinear Hall effect measured from the probe covering the entire ribbon's surfaces.** (a) Berry curvature dipole component $D_{xz}$ at 50K. Red and black dash lines highlighted the BCD values at 0.0 and 0.1 eV. (b) Zoom-in view of the 3.7 nm wide TaIrTe$_4$ nanoribbon's band structure, with red/black dash lines highlighting the center of the transport windows. (c) The in-plane nonlinear Hall effect induced electric field ($E_H$) calculated from voltage measurement from left/right edge as $(V_R - V_L)/W$, where $W$ is the ribbon width. Both forward and reverse bias were applied. The red/blue dash line show the Berry curvature dipole (BCD) induced $E_H$ under 2D bilayer TaIrTe$_4$ structure. (d) The out-of-plane nonlinear Hall effect induced $E_H$ measured from top/bottom surface as $(V_T - V_B)/H$, where $H$ is the ribbon's thickness (~ 1.05 nm). Both in-plane and out-of-plane $E_H$ were extracted in transport windows center at 0.0 and 0.1 eV, respectively.

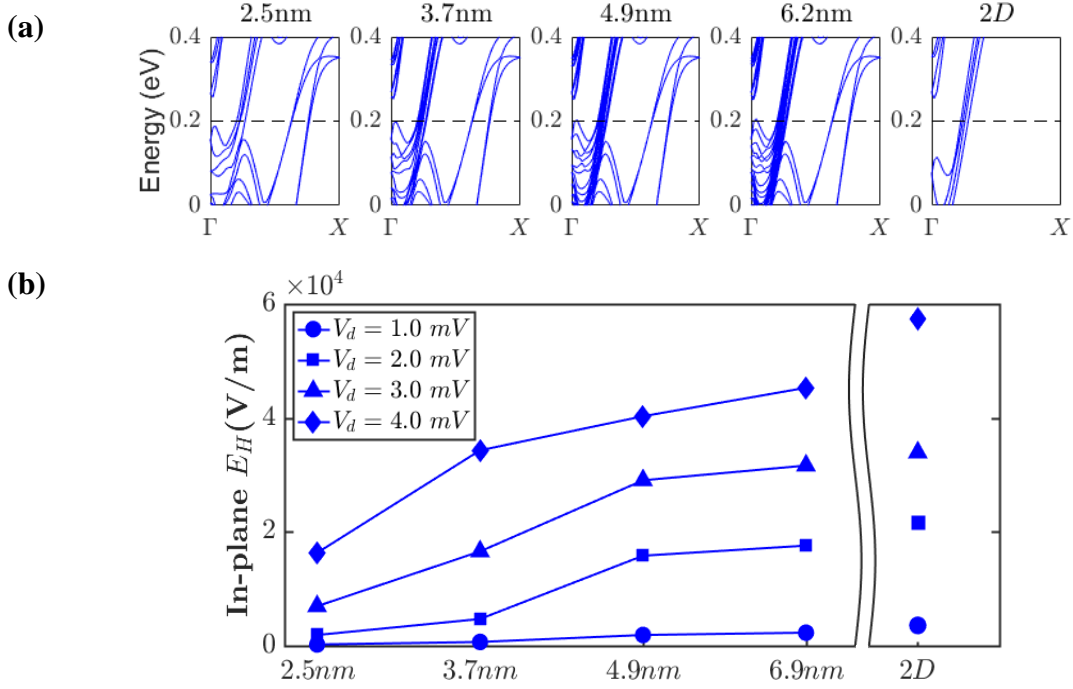

FIG. 3. **Illustration of NLHE's scaling with various ribbon widths.** (a) Zoom-in view of band structures from nanoribbons in transport direction with different width, scaling from 2.5 nm to 6.9 nm. The band structure of 2D counterpart is also included. Black dash lines highlight the center of the transport window at 0.2 eV. (b) In-plane Hall effect induced electric field, scaling with different ribbon width in transport window centered at 0.2 eV. The $E_H$ were extracted from different applied bias $V_d$ from 1 mV to 4 mV under the same longitudinal channel length (~2.3 nm). The Berry curvature dipole induced $E_H$ in two-dimensional structure under the same applied field is labelled as "2D".

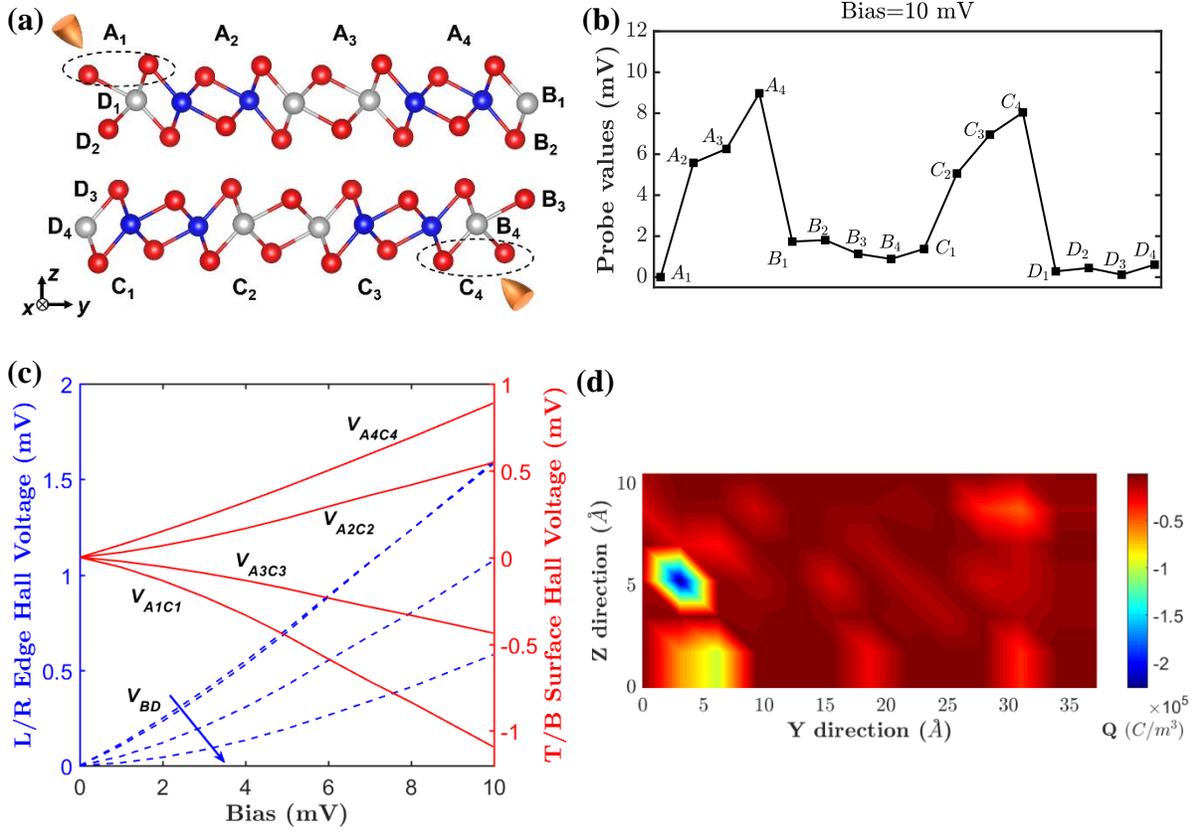

FIG. 4. **Atomic texture of NLHE at the cross-section at $L/2$.** (a) Schematic view of probe scanning along cross-section surface, with probed atoms/atom groups labeled under A,B,C,D. On upper/lower edges the probes are connected to atomic groups formed by two Ta atoms, highlighted by the dash circles. These groups are labelled as $\{A_1…A_4\}$ and $\{C_1…C_4\}$ on upper/lower edge. On left/right edge, the probe resolution changes to single atom, and the probe positions are labelled as $\{B_1…B_4\}$ and $\{D_1…D_4\}$. The transport window is tuned to center at 0.0 eV. (b) Probed potential in each measurement, using the probed potential of atomic groups $A_1$ as reference at the applied bias of 10 mV. (c) Hall Voltage extracted from each individual probe measurement as the function of applied bias. $V_{BD}$ (blue dash lines, to the left $y$ axis) are in-plane components extracted between atoms labelled group B and D. The arrow represents the Hall voltage in the order of $\{V_{B1D1}, V_{B2D2},…,V_{B4D4}\}$. The red lines (to the right axis) represent the out-of-plane Hall voltage with each line labeled with probed positions following the scheme shown in (a). (d) Charge density distribution of the investigated cross-section with the same bias conditions as in (b).

# Supplementary Information

# Real Space Characterization of Nonlinear Hall Effect in Confined Directions

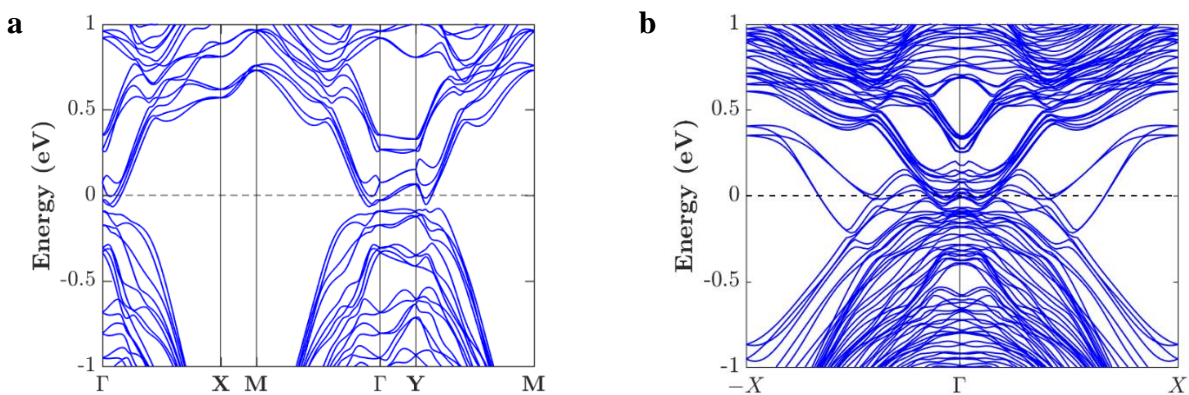

**Figure S1 | Demonstration of bi-layer TaIrTe4 band structures. a**, Band structure of the bi-layer TaIrTe4 2D-sheet, black dash line marking the Fermi level. **b**, Band structure of 3.7 nm-wide bi-layer TaIrTe4 nanoribbon, black dash line marking the Fermi level.

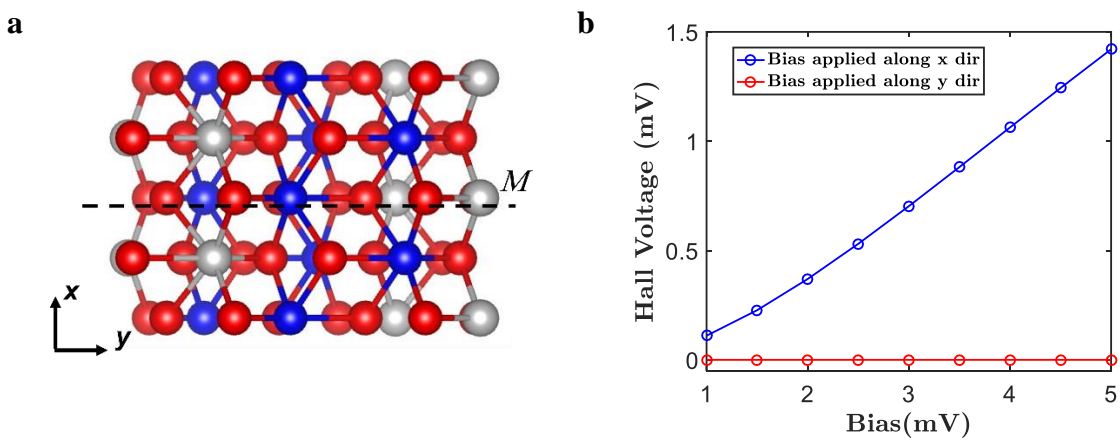

**Figure S2 | Mirror symmetry's constraints on nonlinear Hall voltage. a**, Bilayer TaIrTe4 from birds-eye view, with dash line "*M*" as the mirror axis parallel to *y* direction. **b**, In-plane nonlinear Hall effect measured in global level with bias applied along *x* and *y* direction, respectively. Bias applied along parallel to the mirror axis yielded overall zero Hall voltage in the transverse direction, demonstrating the mirror symmetry constraints on transverse Hall components. In contrast, nonlinear Hall effect could be observed with bias applied along *x* direction.

**Supplementary Note 1 | Probe parameters and probed chemical potential's robustness**

The probe could be implemented in the NEGF framework as self-energy $\Sigma$ through: $\Sigma = \tau g_s \tau^+$, where $g_s$ is the surface Green's function and $\tau$ is the coupling matrix between the contact and the channel [1]. The imaginary part of $g_s$ accounts for the induced level broadening, which could be expressed as: $Im(g_s) \approx \pi * DOS(E_f)$, relating to the density of states of the contact material. By setting the non-zero coupling factor of $\tau$ with certain atom's orbital, the probe is connected to the orbitals of atoms for measurement. Depending on the scale of probing, the self-energy $\Sigma$ could be added to a few atoms or all the atoms on the entire surfaces simultaneously, corresponding to the NLHE probing in local [shown in Fig. 1(c)] or global levels [Fig. 1(a) and (b)]. The metal contact is assumed as an electron injection source characterized by the chemical potential [1], with $\tau$ fixed at $10^{-2}$ eV and the imaginary part of $g_s$ as 1/eV per unit-cell-volume for all probe settings. It is worth noting that the configurations of the probe's self-energy $\Sigma$ are important for the robustness of the probed nonlinear Hall voltage, as the strong interactions of the coupling $\tau$ and broadening might alter the probed Hall voltage or system's symmetry. Therefore, it is necessary to minimize the impact of self-energy $\Sigma$ to obtain robust Hall voltage measurement and ensuring the numerical accuracy for the probed chemical $\mu$. To verify the robustness of the probed nonlinear Hall voltage, various range of self-energy $\Sigma$ are shown in Fig. S3, showing the aforementioned $\tau$ and $g_s$ setting satisfied the robustness of the NLHE for further discussion.

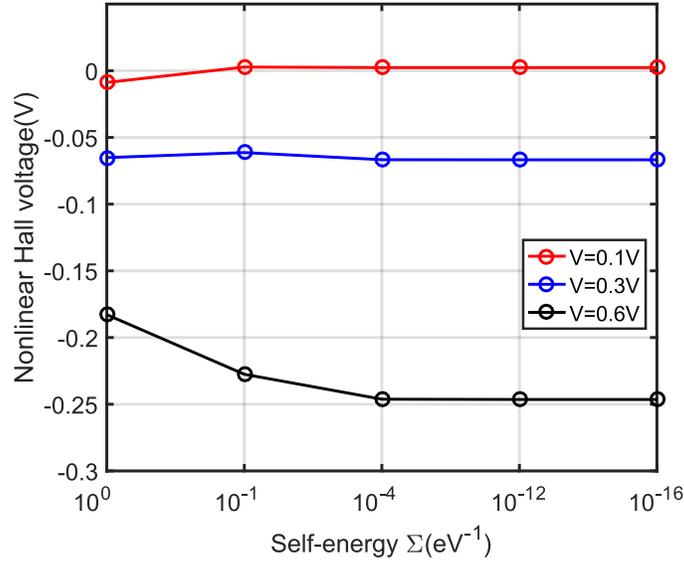

**Figure S3 | Probed nonlinear Hall voltage under different probe settings.** The probed nonlinear voltage at different self-energy $\Sigma$ modeling the probe under different applied bias. To minimize the impact of probe parameters on measured Hall voltage, the value of the self-energy should be minimized. It is noted that the probe setting implemented (at $10^{-4}$ eV$^{-1}$) satisfies the robustness requirement for the voltage measurement.

**Supplementary Note 2 | Overall response of Hall current from NEGF scheme**

It is noted that the non-equilibrium Green's function (NEGF) allows the direct evaluation of overall transverse current through the probe, including both linear, 2$^{nd}$ order response and higher order response. However, the first order response of the probe current vanished when the Hall voltage was extracted by finding the differences of probed chemical potential from opposing edges. This was resulting from the group velocity symmetry in the transverse direction under the time-reversal symmetry constraints [2-3], and it leads to the removal of linear components from the extracted Hall voltage. The removal of the first order components allows the dominance of non-zero 2$^{nd}$ order components derived from broken inversion and mirror symmetry in the overall current response. The numerical accuracy of 2$^{nd}$ order components captured by NEGF could indeed be affected by the higher order components ($3^{rd}$ order and above) under high field conditions [4-6] or suppressed by the mirror symmetry constraints in both in-plane and out-of-plane directions [7]. Nevertheless, these conditions were not presented neither in the simulation configurations, nor the nanoribbon structure of bilayer TaIrTe$_4$.

**Supplementary Note 3 | Different contributions other than potential**

It is worth noting that the besides the non-rigid scattering induced by the potential, the interface between the lead and scattering region (Fig. S4) could also introduce scattering as the electrons enter into the scattering region. This could be of concern as the length of the scattering region configured in the simulation is short (~2.3 nm). In Fig. S5, it demonstrates the dominance of the potential-induced NLHE in the overall NLHE response, and the other non-rigid scattering mechanisms have minor impact. To further mitigate the effect from the interfacial scattering induced components, it is expected that the increment of the scattering region's length could mitigate its impact. Besides, as the leads characterized in the NEGF could represent various material system, the focus of this work is in the NLHE induced by bi-layer TaIrTe$_4$. Therefore, the same material is assumed in the lead region to exclude other factors.

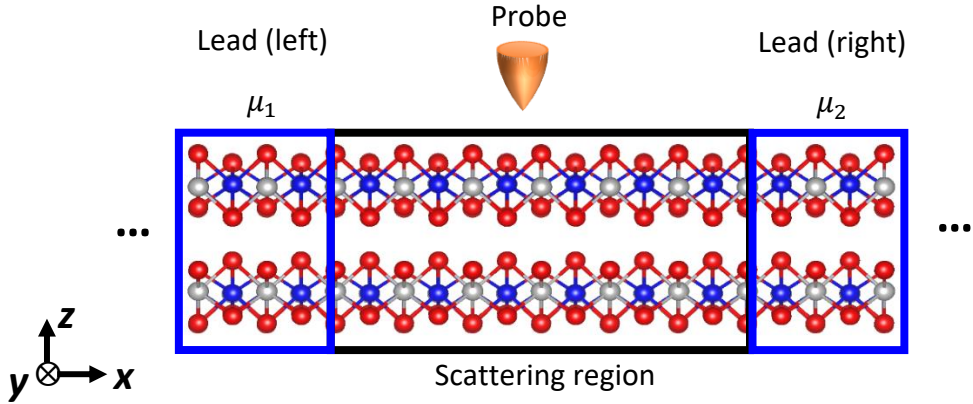

**Figure S4 | Schematic view of the system simulated.** The figure demonstrate a bi-layer TaIrTe$_4$ partitioned by the leads and scattering region. In the longitudinal direction, the TaIrTe$_4$ is assumed to be infinitely long.

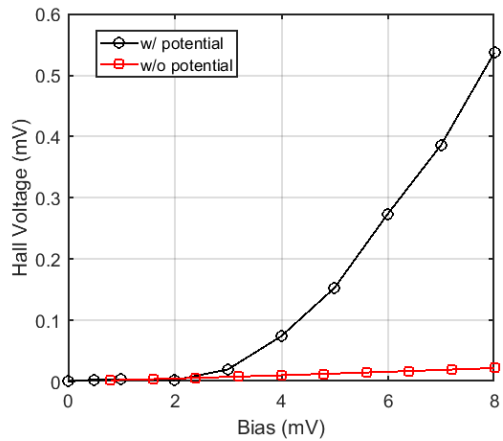

**Figure S5 | Contributions of potential-induced NLHE.** The probed in-plane Hall voltage in global level (the probe covers the entire edge). The contributions from the applied potential dominates the probed nonlinear Hall voltage.